\documentclass[prb,oneside,showpacs,amsmath,amssymb,reprint,superscriptaddress]{revtex4-1}

\usepackage{times}
\usepackage{graphicx}
\usepackage{dcolumn}
\usepackage{bm}
\bibpunct{}{}{,}{s}{}{}

\usepackage{amsmath}    
\usepackage{graphicx}   
\usepackage{verbatim}   
\usepackage{color}      
\usepackage{subfigure}  
\usepackage{braket}
\usepackage{rotating}

\begin{document}

\preprint{Near final version}

\title{Exciton-polaron complexes in pulsed electrically-detected magnetic resonance}
\author{T. L. Keevers}\email{t.keevers@student.unsw.edu.au}
\affiliation{School of Physics, UNSW Australia, Sydney NSW 2052, Australia}
\author{W. J. Baker}
\affiliation{School of Physics, UNSW Australia, Sydney NSW 2052, Australia}
\affiliation{Centre for Quantum Computation and Communication Technology, UNSW Australia}
\author{D. R. McCamey}
\affiliation{School of Physics, UNSW Australia, Sydney NSW 2052, Australia}

\date{\today}

\begin{abstract}
Several microscopic pathways have been proposed to explain the large magnetic effects observed in organic semiconductors, but identifying and characterising which microscopic process actually influences the overall magnetic field response is challenging. Pulsed electrically-detected magnetic resonance provides an ideal platform for this task as it intrinsically monitors the charge carriers of interest and provides dynamical information which is inaccessible through conventional magnetoconductance measurements. Here we develop a general time domain theory to describe the spin-dependent reaction of exciton-charge complexes following the coherent manipulation of paramagnetic centers through electron spin resonance. A general Hamiltonian is treated, and it is shown that the transition frequencies and resonance positions of the exciton-polaron complex can be used to estimate inter-species coupling. This work also provides a general formalism for analysing multi-pulse experiments which can be used to extract relaxation and transport rates. 
\end{abstract}

\maketitle

\section{Introduction}

Organic devices with non-magnetic electrodes may display large magnetoresistances at room temperature\citep{francis2004large}. Magnetic fields of 10 mT or less are sufficient to induce large changes in sample conductivity\citep{mermer2005large}, with reports of changes exceeding 2000 percent in one-dimensional wires\citep{mahato2013ultrahigh}. There is interest in harnessing these field-induced effects to enhance technologies including solar cells\citep{hoppe2004organic}, organic light-emitting diodes\citep{dodabalapur1997organic}, and cheap magnetic sensors\cite{baker2012robust}. This requires sophisticated methods for characterizing and engineering of the underlying spin processes. 

Several detailed models\citep{prigodin2006anomalous,desai2007magnetoresistance,bobbert2007bipolaron} have been developed to explain the ubiquitous magnetoconductance effects observed in organic devices. In recent years there has been significant progress in distinguishing potential spin processes through lineshape and voltage fingerprints\citep{janssen2013tuning}, however, these types of quasi-static measurements and models are fundamentally limited: Dynamical rates and energetic couplings are obscured by the effects of spin ensemble averaging and the intrinsically incoherent nature of the measurement. Coherent spectroscopic techniques, such as pulsed electron paramagnetic resonance (EPR), provide a way to independently identify and quantify the underlying physical processes because they allow sub-ensemble of spins to be selectively addressed and manipulated within the coherence time\citep{boehme2013challenges}. 

Pulsed Electrically Detected Magnetic Resonance\citep{Boe03} (pEDMR) in particular provides a strong spectroscopic probe with which to investigate these processes, as different spin processes can readily be disentangled\citep{boehme2013challenges,keevers2013agnostic} and their impact on device operation directly observed from the transient dynamics. Pulsed methods are able to access coherent time domain information and by adjusting the applied pulse sequence different aspects of the spin ensemble can be probed, allowing a holistic picture of the spin physics to be developed and rigorously tested\citep{schweiger2001principles}. Past studies have primarily focused on pristine Poly[2-methoxy-5-(2-ethylhexyloxy)-1,4-phenylenevinylene] (MEH-PPV) devices in which polaron pairs are the dominant spin-reaction pathway, and a detailed picture of the relevant energy and time scales in these devices has been developed\citep{mccamey2010hyperfine,baker2012slow}. Dissociation, recombination, and intersystem crossing rates have been characterized\citep{mccamey2008spin,mccamey2010spin}, as well as the strength of the electro-nuclear coupling and spin-spin interactions within the pair\citep{malissa2014room}. Recent improvements in pEDMR measurement methods have substantially improved signal-to-noise\citep{hoehne2012lock}, and a variety of standard one and two-dimensional experiments from pEPR\citep{schweiger2001principles} are now technically feasible. In contrast, energy and time scales cannot be effectively separated in quasi-static magnetoconductance measurements; Ultra-small field effects are consistent with either a weak exchange interaction\citep{nguyen2010magnetoconductance} or slow carrier hopping\cite{schellekens2011microscopic}. Differentiation between these two scenarios is possible with pulsed measurements\citep{mccamey2010hyperfine,baker2012slow}, however no comparable fingerprint is known for the static case. 

An extensive body of literature has been developed for describing the time domain response of polaron pairs in magnetic resonance\citep{rajevac2006transport,gliesche2008effect,limes2013numerical}, while comparatively little is known in relation to possible three-particle processes. These are expected to include triplet-exciton polaron quenching or the dynamics of a bipolaron stabilized by a counter-ion in organic systems\citep{shinar2012optically}. Other three-particle processes also include spin-dependent Auger recombination in nanocrystals\citep{van2014coherent}, trion states in silicon\citep{belolipetskiy2014trions}, and electrons interacting with a common nuclear bath\citep{weber1997transient}.

In this work we develop a general formalism for describing the spin-dependent response of three-particle complexes in pulsed magnetic resonance, thus providing the theoretical foundations for the quantitative characterization of the time and energy scales present in these systems. 

There are a number of different mechanisms invoked in the literature to describe the triplet-exciton polaron (TEP) quenching process. For instance, Desai et al\citep{desai2007magnetoresistance} described a mechanism in which free polarons scatter from trapped triplet-excitons, causing the exciton to cross form the triplet to singlet manifold. 

\begin{equation}
T_1 + D_{\pm 1/2} \leftrightarrow (T_1 ... D_{\pm 1/2}) \rightarrow D_{\pm 1/2} + S_0^*
\end{equation}

Where $T_1$ is an excited triplet exciton, D is a free charge carrier and S is a singlet exciton. The reaction rate was assumed to be spin-dependent, although no microscopic mechanism was provided. In contrast, Koopmans and co-workers describe a quenching process which depends on the overall quartet or doublet content of the pair and may increase or decrease the polaron's mobility\citep{schellekens2011microscopic}. Other implementations include a site-blocking mechanism\citep{song2010effect} and an exciton-polaron exchange process\citep{song2010effect}.

Similarly, there is substantial variation in the types of interactions assumed within the three-spin complex and the conditions under which they will be observed. The triplet-exciton polaron (TEP) process has been discussed in terms of weakly-interacting precursor pairs\citep{schellekens2011microscopic}, strongly-bound trion states\citep{cox2013traps}, and even bipolaron-counter ion complexes\citep{shinar2012optically}. Each of these scenarios will produce a unique magnetic field-dependence and differentiation is an important step towards purposeful tuning of spintronic devices. The conditions under which the TEP mechanism will appear are also debated: High magnetic-field features from magnetoconductance measurements indicate that the triplet process is important at room temperature\citep{janssen2013tuning}, in contradiction with the sharp decrease in triplet lifetimes observed at high temperatures\citep{baldo2000transient} and in earlier pEDMR experiments\citep{baker2011differentiation}. 

In this paper we develop a time domain theory of an interacting spin-1+spin-1/2 pair which can react in proportion to its overall doublet (S=1/2) content, analogous to the theory of Ern and Merrifield\citep{ern1968magnetic}. The transition frequencies and resonance positions of an exciton-polaron complex are extracted from the canonical Hamiltonian, providing an analytic description of the transition between a weakly spin-interacting exciton-polaron complex and a strongly-bound trion state. This result can be used to analyze pEDMR experiments on organic devices, allowing the intra-pair spin-spin coupling to be extracted, as demonstrated in related work\cite{baker2014tripletexpt}. In this paper we concentrate on transitions following a single pulse, however the formalism and can be applied more generally to multi-pulse sequences such as the Hahn echo\citep{Huebl2008,Bak12}. As such, it provides the foundations for quantitative investigation of the triplet-exciton polaron quenching process with pEDMR.

\section{General features}

Long-lived triplet excitons can interact with trapped polarons to form complexes, which may undergo a charge-reaction process which releases the trapped polaron and alters the free charge carrier density. Although a similar process is possible with singlet excitons, their density will be much lower\citep{shinar2012optically} and we therefore ignore their contribution in this work. The spin-independent formation of exciton-charge complexes is likely to dominated by the reorganization energy associated with the shared lattice distortion\citep{stafstrom2010electron} due to the weak spin-interaction between the exciton and polaron. 

Following its formation, an exciton-charge complex may i) dissociate back into its constituents or ii) recombine if it is in the doublet (spin-1/2) manifold, which will release the trapped polaron into the valence or conduction band, increasing the charge carrier mobility (see Fig. \ref{fig:Model}a). The exact states involved in the reaction are unclear, however the weak spin-coupling seen experimentally \citep{baker2011differentiation} suggests that a strongly-bound trion state is formed prior to recombination.

\begin{equation}
[Exciton + Polaron] \rightarrow Exciton + Polaron
\end{equation}

\begin{equation}
[Exciton + Polaron] \rightarrow [Trion] \rightarrow S_0 + Polaron^*
\end{equation}


Spin resonance causes a mixing of the doublet and quartet manifolds, causing a net change in the recombination rate and average charge carrier mobility (conductivity). As in the two spin polaron-pair case\citep{Boe03} the second order contributions due to change in the electron and hole bath concentrations and mobilities are negligible, thus the expected changes in mobility are given by

\begin{multline}
\\
\Delta \mu_e(t) = \tau_L (R_e(t)-R_{e,eq}) \\
\Delta \mu_h(t) = \tau_L (R_h(t)-R_{h,eq}) \\
\end{multline}
Where $\tau_L$ is the carrier lifetime, $n_{e,h}$ are the number of free electrons and hole, and $R(t)$ and $R_{eq}$ are the time-dependent and equilibrium recombination rates. 

The increase in charge carrier mobility from the charge reaction may yield either a current enhancement or quenching, depending on whether the majority or minority carriers are released. This counterintuitive behaviour arises due to the interplay between coulombic attraction and carrier transport\citep{bloom2009sign}. In general, we expect the charge reaction to release the majority carrier and provide a current enhancement. In the unipolar regime, which we treat as electron-rich, the change in current will be proportional to the number of released carrier. 
\begin{equation}
\Delta \sigma(t) = e \Delta \bar{\mu}_e(t) n_e
\end{equation}

For a general description of the spin dynamics, we need to include the interaction of the spin ensemble with the environment. This includes charge carrier generation, dissociation, recombination and relaxation. These influences may be included through the stochastic Liouville equation\citep{schellekens2011microscopic}
\begin{equation}
\frac{\partial \rho}{\partial t} = \frac{i}{\hbar} [\rho,H] + S[\rho] + R\{\rho - \rho_0\}
\end{equation}

Where $\rho$ is the density matrix and represents the occupation probability of each state, H is the coherent spin Hamiltonian and describes the exciton-polaron spin complex, S is a stochastic operator and represents the generation and loss of charge carriers, and R is the Redfield relaxation matrix and incorporates incoherent effects such as interactions with the surrounding phonon bath.

\subsection{Hamiltonian}

\begin{figure*} [t]
\centerline{\includegraphics[width=16cm]{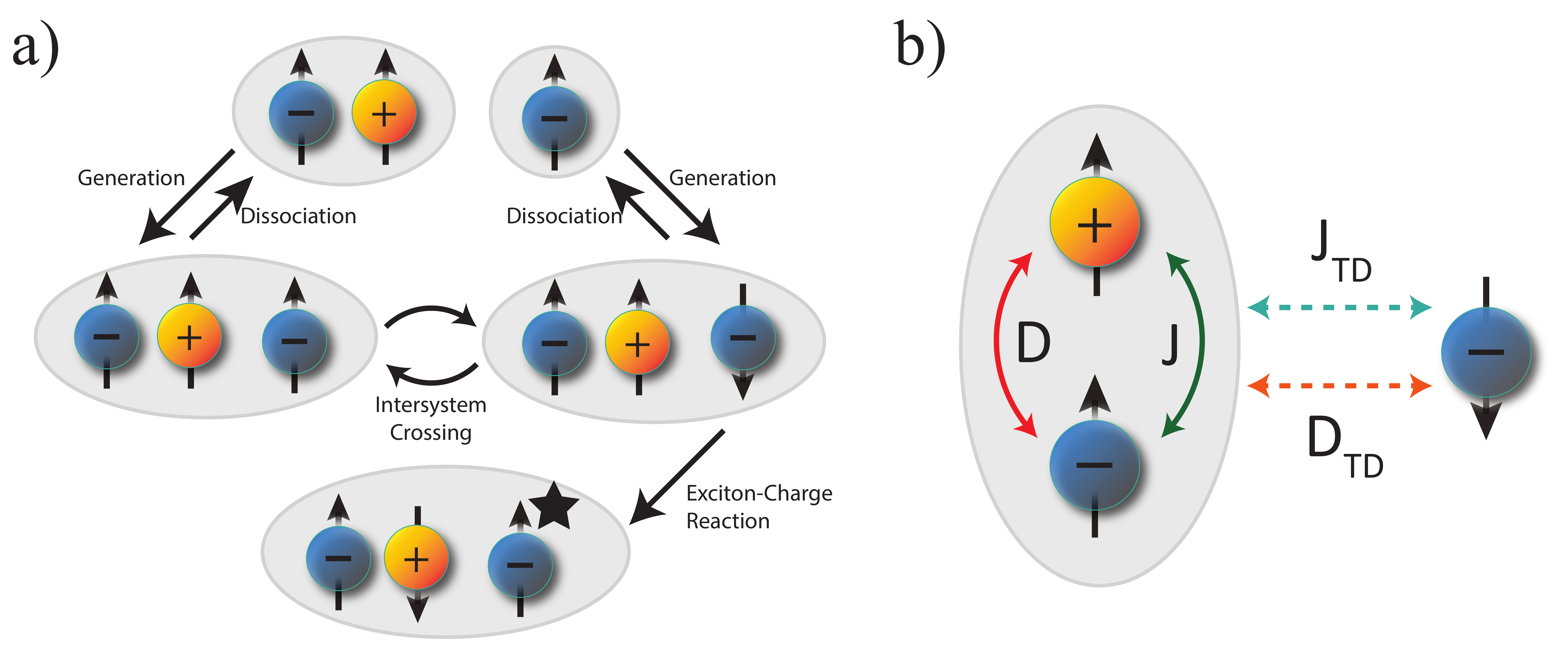}}
\caption{a) The general picture of the triplet exciton-polaron quenching mechanism. Exciton-polaron complexes are formed spin-independently due to the reorganization energy associated with a shared nuclear distortion. A complex may spin-independently dissociate back in a separate exciton and polaron or may undergo a exciton-charge reaction, possibly due to weak exciton-polaron exchange. Intersystem crossing will mix the doublet and quartet states prior to recombination or dissociation. b) A diagramatic representation of the Hamiltonian for an exciton-polaron complex with arbitrary coupling. The exciton will have internal exchange and dipolar coupling, and there may also be exchange and dipolar coupling between the exciton and the polaron.}
\label{fig:Model}
\end{figure*}

The spin Hamiltonian describes the coupling within the exciton-polaron complex and its interactions with the surrounding nuclear environment. As it completely encapsulates the coherent spin dynamics it can be used to derive the resonance positions and transition frequencies which allow the mechanism and coupling strengths to be identified. Both the exciton and polaron couple to the nearby nuclei through the hyperfine interaction and to the external magnetic field through Zeeman splitting. There is also dipolar coupling within the exciton, and dipolar and exchange coupling between the exciton and polaron (see Fig. \ref{fig:Model}b). The exciton has a large internal exchange (J $\approx$ 0.7 eV) which separates the singlet and triplet levels. As the internal exchange is significantly larger than all other energy scales in a typical experimental regime, the singlet level becomes decoupled and as such the exchange interaction can be excluded from the effective Hamiltonian.

\begin{multline}\label{HamText}
H = \underbrace{\beta_e B \cdot g_T \cdot S_T}_{\text{Exciton Zeeman}} + \underbrace{\beta_D B \cdot g_D \cdot S_D}_{\text{Polaron Zeeman}} + \\
\sum_i \underbrace{I_{Ti} \cdot A_{Ti} \cdot I_{Ti}}_{\text{Exciton Hyperfine}} + \sum_i \underbrace{I_{Di} \cdot A_{Di} \cdot I_{Di}}_{\text{Polaron Hyperfine}} + \\
\underbrace{S_T \cdot D_T \cdot S_T}_{\text{Exciton Dipolar}} + \underbrace{J_{TD} S_T \cdot S_D}_{\text{Exciton-Polaron Exchange}} \\
+ \underbrace{S_T \cdot D_{TD} \cdot S_D}_{\text{Exciton-Polaron Dipolar}} \\
\end{multline}

An expanded version of this matrix is given in Appendix \ref{HamiltonianAppendix}. The commonly used high field approximations are made, which include an isotropic g-factor, isotropic hyperfine, and only retaining the secular terms of the dipolar coupling\cite{atherton1993principles}.

Starting from the triplet-doublet (product) basis we can rotate the Hamiltonian to the energy eigenbasis using the unitary transform
  \begin{equation}
H_{energy} = U^{\dag} H U
\end{equation}
where
    \begin{equation}\label{eq:unitarytrans}
    U = 
      \begin{bmatrix}
1 & 0 & 0 & 0 & 0 & 0 \\
0 & \cos{\theta} & 0 & -\sin{\theta} & 0 & 0 \\
0 & 0 & \cos{\phi} & 0 & \sin{\phi} & 0 \\
0 & \sin{\theta} & 0 & \cos{\theta} & 0 & 0 \\
0 & 0 & -\sin{\phi} & 0 & \cos{\phi} & 0 \\
0 & 0 & 0 & 0 & 0 & 1 
      \end{bmatrix}
    \end{equation}
    
 with

\begin{equation}
\cot{2 \theta} = \frac{D_T - \frac{D_{TD}}{3} + \frac{J_{TD}}{2} - \omega_D + \omega_T}{\frac{\sqrt{2} D_{TD}}{3} + \sqrt{2} J_{TD}}
\end{equation}

and 
\begin{equation}
\cot{2 \phi} = \frac{D_T - \frac{D_{TD}}{3} + \frac{J_{TD}}{2} + \omega_D - \omega_T}{\frac{\sqrt{2} D_{TD}}{3} + \sqrt{2} J_{TD}}
\end{equation}

This unitary transform preserves the physics of the Hamiltonian, while simplifying analytic and numerical calculations. For instance, the transition frequencies simply correspond to the off-diagonal matrix elements. Importantly, the angles $\theta$ and $\phi$ represent the degree of coupling between the exciton and polaron. Setting $\theta,\phi$ = 0 corresponds to two independent particles and $\theta,\phi = \cos^{-1}{\sqrt{\frac{2}{3}}}$ (the maximum rotation angle possible) corresponds to the formation of a spin-$\frac{3}{2}$ trion. Due to the convenient measure of spin-spin coupling that $\theta$ and $\phi$ provide we will refer to them throughout the text as the $\textit{spin mixing angles}$. 

The three natural basis sets for describing an exciton-polaron complex are the triplet-doublet (product) basis, the quartet-doublet basis and the energy eigenbasis. The transformation between the triplet-doublet and quartet-doublet bases is given by the Clebsch-Gordan coefficients.

\begin{eqnarray}
\ket{T_+ \uparrow} & = & Q \nonumber \\
\ket{T_+ \downarrow} & = & \sqrt{\frac{1}{3}} Q + \sqrt{\frac{2}{3}} D \nonumber \\
\ket{T_0 \uparrow} & = & \sqrt{\frac{2}{3}} Q - \sqrt{\frac{1}{3}} D \nonumber \\
\ket{T_0 \downarrow} & = & \sqrt{\frac{1}{3}} Q + \sqrt{\frac{2}{3}} D \nonumber \\
\ket{T_- \uparrow} & = & -\sqrt{\frac{2}{3}} Q + \sqrt{\frac{1}{3}} D \nonumber \\
\ket{T_- \downarrow} & =  & -Q \nonumber \\
\end{eqnarray}

It is important to note that the term `doublet' refers to a state with S=1/2, which may either be a single spin-1/2 particle or the $\ket{S,m_s} = \ket{\frac{1}{2},\pm \frac{1}{2}}$ trion states. The transformation from the triplet-doublet to the energy eigenbasis is given by

\begin{equation}
\rho_{energy} = U^{\dag} \rho
\end{equation}

with the triplet-doublet states given by
    \begin{equation}
    \rho = 
      \begin{bmatrix}
\ket{T_+\uparrow} \\
\cos{\theta} \ket{T_0\uparrow} + \sin{\theta} \ket{T_+\downarrow} \\
\cos{\phi} \ket{T_-\uparrow} - \cos{\phi} \ket{T_0\downarrow} \\
\cos{\theta} \ket{T_+\downarrow} - \sin{\theta} \ket{T_0\uparrow} \\
\cos{\phi} \ket{T_0\downarrow} + \sin{\phi} \ket{T_-\uparrow} \\
\ket{T_-\downarrow}
      \end{bmatrix}
    \end{equation}
    
It is conceptually and computationally expedient to split the Hamiltonian into a large static component ($H_0$) comprising of spin-spin coupling and the static Zeeman splitting and a small, time-dependent driving term ($H_1$) which corresponds to the time-dependent Zeeman splitting.

\begin{equation}
H = H_0 + H_1
\end{equation}

\begin{equation}
H_1 = 2 B_1 (e^{-i \omega t} + e^{i \omega t}) (S_1^x + S_2^x)
\end{equation}

At high field the counter-propagating component will produce no first-order effects, so we can move from the energy eigenbasis to the rotating frame and perform the rotating wave approximation.

\begin{equation}
H_{rot} = U_{rot}^\dag H U_{rot} - U_{rot}^{\dag} \frac{\partial H}{\partial t} U_{rot}
\end{equation}

\begin{equation}
U_{rot}(t) = e^{-i B_0 (S_1^z + S_2^z) t}
\end{equation}
    
In the rotating eigenbasis we have the $H_0$ energy levels on the diagonal entries and the perturbative spin mixing $H_1$ terms on the off-diagonal elements. 

    \begin{equation}\label{mat:Ham}
    H_{rot} = 
      \begin{bmatrix}
E_1 & B_{12} & 0 & B_{14} & 0 & 0 \\
B_{21} & E_2 & B_{23} & 0 & B_{25} & 0 \\
0 & B_{32} & E_3 & B_{34} & 0 & B_{36} \\
B_{41} & 0 & B_{43} & E_4 & B_{45} & 0 \\
0 & B_{52} & 0 & B_{54} & E_5 & B_{56} \\
0 & 0 & B_{63} & 0 & B_{65} & E_6
      \end{bmatrix}
    \end{equation}

The dressed rotating basis states will have the same form as the bare energy eigenbasis states, and for most situations they may be treated equivalently\cite{schweiger2001principles}.      
    
\begin{figure*} [t]
\centerline{\includegraphics[width=16cm]{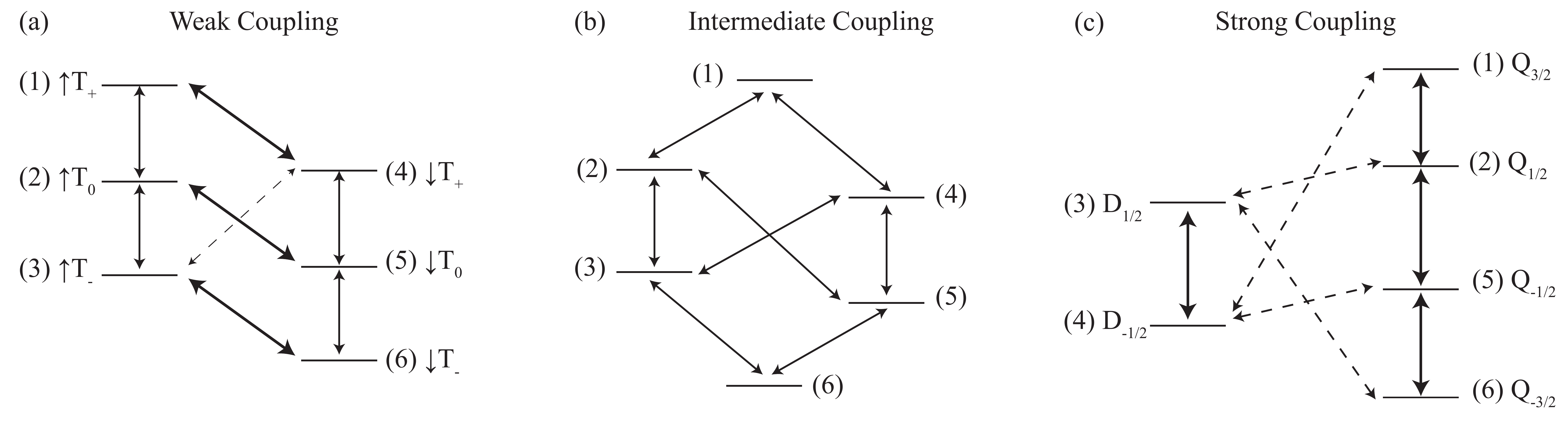}}
\caption{The behaviour of an exciton-polaron complex depends strongly on the strength of the exciton-polaron coupling. Solid lines represent strongly-driven transitions and dashed lines represent transitions which become forbidden in the weak or strong coupling limits. a) In the weak regime the basis states are close to the product states and the transitions are the $\uparrow \leftrightarrow \downarrow$ polaron transition, with a Rabi frequency of $\gamma B_1$, and the $T_+ \leftrightarrow T_0$, $T_0 \leftrightarrow T_-$ exciton transitions with a Rabi frequency of $\sqrt{2} \gamma B_1$. The $\downarrow T_+ \leftrightarrow \uparrow T_-$ transition becomes weakly-allowed in the presence of non-zero coupling. b) In the intermediate coupling regimes the basis states are an admixture of the product and quartet-doublet states and up to eight transitions are visible with Rabi frequencies between 0 and $\sqrt{3} \gamma B_1$. c) In the strongly-coupled regime quartet and doublets are formed. Strong driving occurs within the doublet and quartet manifolds and driving between the manifolds becomes arbitrarily slow, leading to a vanishing signal. The Rabi frequencies of these transitions are $\sqrt{3} \gamma B_1$ and 2$\gamma B_1$.}
\label{fig:CouplingRegimes}
\end{figure*}

\subsection{Limiting behaviour of the Hamiltonian}

While numerical calculations provide a general method for calculating the exciton-polaron dynamics from the spin Hamiltonian, this approach limit the physics insight which we can obtain. To understand the general properties of the system we now look at a number of limiting cases.

\subsubsection{Weak coupling: $\theta,\phi \rightarrow$ 0}

When the exciton-polaron coupling is extremely weak ($\theta,\phi \rightarrow$ 0) the spins will be driven independently, and the eigenbasis is given trivially by the product states. The Hamiltonian for this scenario is

    \begin{equation}
    H = 
      \begin{bmatrix}
E_{T_+ \uparrow} & \frac{B_1}{\sqrt{2}} & 0 & \frac{B_1}{2} & 0 & 0 \\
\frac{B_1}{\sqrt{2}} & E_{T_0 \uparrow} & \frac{B_1}{\sqrt{2}} & 0 & \frac{B_1}{2} & 0 \\
0 & \frac{B_1}{\sqrt{2}} & E_{T_- \uparrow} & 0 & 0 & \frac{B_1}{2} \\
\frac{B_1}{2} & 0 & 0 & E_{T_+ \downarrow} & \frac{B_1}{\sqrt{2}} & 0 \\
0 & \frac{B_1}{2} & 0 & \frac{B_1}{\sqrt{2}} & E_{T_0 \downarrow} & \frac{B_1}{\sqrt{2}} \\
0 & 0 & \frac{B_1}{2} & 0 & \frac{B_1}{\sqrt{2}} & E_{T_- \downarrow}
      \end{bmatrix}
    \end{equation}
    
\begin{equation}
    \rho = 
      \begin{bmatrix}
T_+\uparrow \\
T_0\uparrow \\
T_-\uparrow \\
T_+\downarrow \\
T_0\downarrow \\
T_-\downarrow
      \end{bmatrix}
    \end{equation}

The polaron resonances occur at a frequency of $\omega_s + \omega_D$ with a Rabi frequency of $\gamma B_1$ and the exciton will have resonate at frequencies of $\omega_s + \omega_T \pm D$ with nutation frequencies of $\sqrt{2} \gamma B_1$. This is shown graphically in Fig. \ref{fig:CouplingRegimes}a). There are three polaron transitions which occur independently of the exciton state, and likewise four possible exciton transitions which are independent of the polaron, shown by the solid lines. The single dashed line represents an $\Delta m_s$ = 1 transition between the $\ket{\uparrow T_-}$ and $\ket{\downarrow T_+}$ states which becomes weakly-allowed in the presence of non-zero exciton-polaron coupling. Eventually the spin-spin coupling becomes large compared to the hyperfine interaction and the transitions can no longer be described as excitonic or polaronic transitions, depicted in Fig. \ref{fig:CouplingRegimes}b.
  
\subsubsection{Strong coupling: $\theta,\phi \rightarrow \cos^{-1}{(\sqrt{\frac{2}{3}}})$} 

When the exciton-polaron coupling is extremely strong ($\theta,\phi \rightarrow \cos^{-1}{(\sqrt{\frac{2}{3}})}$) a trion state will be formed in the quartet-doublet basis. The exciton and polaron can no longer be considered separate entities and they will always nutate together. The Hamiltonian is given by

    \begin{equation}
		H =
      \begin{bmatrix}
E_{Q=\frac{3}{2}} & \frac{\sqrt{3}}{2} B_1 & 0 & 0 & 0 & 0 \\
\frac{\sqrt{3}}{2} B_1 & E_{Q=\frac{1}{2}} & 0 & 0 & 2 \gamma B_1 & 0 \\
0 & 0 & E_{D=\frac{1}{2}} & \gamma B_1 & 0 & 0 \\
0 & 0 & \gamma B_1 & E_{Q=-\frac{1}{2}} & 0 & 0 \\
0 & 2 \gamma B_1 & 0 & 0 & E_{Q=-\frac{1}{2}} & \frac{\sqrt{3}}{2} B_1 \\
0 & 0 & 0 & 0 & \frac{\sqrt{3}}{2} B_1 & E_{Q=-\frac{3}{2}}
      \end{bmatrix}
    \end{equation}
    
with the corresponding eigenstates

\begin{equation}
    \rho = 
      \begin{bmatrix}
Q_{\frac{3}{2}} \\ 
Q_{\frac{1}{2}} \\
D_{\frac{1}{2}} \\
D_{-\frac{1}{2}} \\
Q_{-\frac{1}{2}} \\
Q_{-\frac{3}{2}}
      \end{bmatrix}
    \end{equation}

The mixing will be between the $m_s = \pm 3/2$ and $m_s = \pm 1/2$ quartet manifolds with nutation frequencies of $\sqrt{3} \gamma B_1$, and mixing within the $m_s = \pm 1/2$ doublet and quartet manifolds with Rabi frequencies of 2$\gamma B_1$, as shown in Fig. \ref{fig:CouplingRegimes}c. The four solid lines represent the strongly-allowed transitions of the trion state and the dashed lines represent quartet-doublet transitions which vanish in the strong coupling regime. Counter-intuitively the $m_s = \pm 3/2$ quartet states actually couple to the $m_s = \mp 1/2$ doublet states, which is nominally an $m_s$ = 2 transition. Analogous to the exciton half-field transition, it becomes weakly-allowed due to off-diagonal mixing in the quartet-doublet basis. Because the strongly-allowed transitions induce no overall change in the doublet or quartet content of the complex the trion states will be electrically undetectable. 

In this paper we analyze the transition between the uncoupled and trionic regimes.

\subsection{Complex relaxation and recombination}

The recombination and dissociation rates of each eigenstate will be the weighted average of its quartet and doublet content. In analogy with the polaron pair\citep{Boe03} case we can form new rates for the admixture states. 

\begin{equation}
r_i = r_D |\braket{i|D}|^2 + r_Q |\braket{i|Q}|^2 
\end{equation}

\begin{equation}
d_i = d_D |\braket{i|D}|^2 + d_Q |\braket{i|Q}|^2 
\end{equation}

The spin loss rate is the sum of the dissociation and recombination rates.

\begin{equation}
\gamma_i = r_i + d_i
\end{equation}

A spin-dependent depopulation of the doublet-state has been observed in other systems and is consistent with simple energetic arguments and spin-selection rules\citep{asano1999intramolecular}. In the weak coupling limit ($\theta,\phi \rightarrow$ 0) the $T_- \uparrow,T_+ \downarrow$ states have $\frac{2}{3}$ doublet-content, and the $T_0$ states each have $\frac{1}{3}$ doublet-content.

\begin{equation}
    r = 
      \begin{bmatrix}
      r_Q \\
\frac{1}{3} r_D + \frac{1}{3} r_Q \\
\frac{2}{3} r_D + \frac{2}{3} r_Q \\
\frac{2}{3} r_D + \frac{2}{3} r_Q \\
\frac{1}{3} r_D + \frac{1}{3} r_Q \\
r_Q
      \end{bmatrix}
    \end{equation}
    
In the opposite, strong coupling limit ($\theta,\phi \rightarrow \cos^{-1}{\sqrt{\frac{2}{3}}}$) the states become pure quartet and doublet states. 
   
   \begin{equation}
    r = 
      \begin{bmatrix}
r_Q \\
r_Q \\
r_D \\
r_D \\
r_Q \\
r_Q
      \end{bmatrix}
    \end{equation}

Between these two limits there will be a monotonic change in the recombination rates as the product basis morphs into the quartet-doublet basis.

\subsection{Complex formation and polarization}

We can't discuss the transitions of an exciton-polaron complex without discussing their formation and steady state polarization: The presence of an electrically-detectable signal intrinsically requires a net transfer of the spin population. 

Exciton-polaron complexes are believed to form spin-independently. A spin-dependent formation has been proposed in the context of polaron pairs, and a similar effect for exciton-polarons will change the intensity of the transitions, but leave the qualitative conclusions of this study unaffected. Regardless, the generation of each state needs to be weighted by a Boltzmann factor to account for the thermal distribution.

\begin{equation}
g_i = g \frac{\exp{(-\frac{E_i}{kT}})}{\sum_i \exp{(-\frac{E_i}{kT}})}
\end{equation}

At high temperatures polarization effects can be ignored and a uniform generation rate can be used. The steady state occupation of each state corresponds to when the pair generation and loss rates are equal.

\begin{equation}
\rho_i = \frac{g_i}{r_i+d_i}
\end{equation}

When doublet states are lost much faster than quartet states there will be a strong polarization of pure quartet states.

\begin{equation}
\rho_0 = \frac{1}{2} (T_+ \uparrow + T_- \downarrow)
\end{equation}

In organic devices the polarization effects arise primarily from the different rates of charge carrier loss\citep{Boe03}. In contrast, porphyrin system may generate a polarization within the triplet-doublet pairs from the intersystem crossing of light-induced singlet-doublet states\citep{kandrashkin2003light,van2002light,kobori1998exchange}.

\subsection{Transient readout}

The real-time detection of sample conductivity changes during resonant excitation is technologically challenging\citep{hoehne2013real}. Consequently, experiments are usually performed using a pump-probe approach, which involves monitoring the (slow) response of the system following the (rapid) magnetic resonant excitation\citep{Boe03}.

In the triplet-exciton polaron quenching mechanism recombination will trap or detrap the polaron and change the average charge carrier mobility. This manifests macroscopically as a change in the sample conductivity. We can therefore relate the recombination rate or overall doublet content of the spin ensemble in our model to an experimental change in conductivity. Larmor beating will dephase the exciton and polaron on a time scale much faster than the transient readout, and we can therefore treat the spin ensemble incoherently using rate equations\citep{gliesche2008effect}

\begin{equation}
R(\tau) = \sum_{i} \delta \rho_{ii} (\tau) r_i
\end{equation}
Where $\delta \rho_{ii}$ is the change in occupation probability and $r_i$ the recombination rate of each state. The overall change in current induced by each transition will depend on the precise choice of integration window and dissociation, recombination, and intersystem crossing rates. In this work, we are concerned with the resonance positions and transition frequencies, which are unaffected by these incoherent rates. Nonetheless, we sketch how quantitative predictions may be made by discussing the minimalist scenario in which intersystem crossing and recombination are slow compared to dissociation, which produces an observable which is proportional to the doublet content of the occupied states. We also assume that all recombination events are recorded.

\begin{equation}
Q = \sum_i \int_0^T r_i \exp{(-(r_i+d_i)t)} dt \approx \sum_i \frac{r_i}{d_i} \approx \sum_i |\braket{D|\phi_i}|^2
\end{equation}
Where $\phi_i$ is the i-th eigenvector and D is the doublet manifold. An unknown scaling factor is required to relate Q, the overall microscopic change in charge carrier mobility, to a real change in sample conductivity.

\section{Transition analysis}

In the standard implementation of a pEDMR experiment the transition frequencies are mapped as a function of energy by sweeping the magnetic field with a constant excitation frequency and measuring the change in charge as a function of the microwave pulse length. This may result in a number of different resonances each with a particular transition frequency. These serve as a `Rabi fingerprint' which can be used to identify and quantify the types of coupling present. We now calculate these `fingerprints' directly from the spin Hamiltonian.

\subsection{Resonance position}

The photon energy required to resonantly drive a transition is equal to the energy difference between the two eigenstates of interest. The energies of the Hamiltonian described by equation (\ref{eq:Energies}) which are used to calculate the resonance positions are given in Table \ref{Tab:ResPosition}.

In the weak coupling limit the three polaron transitions will occur at the Larmor precession frequency and the four exciton transitions will be paired and will be offset the intra-exciton dipolar coupling. The multiplicity of each transition is due to possible states of its partner, so the exciton has identical transitions whether the polaron is in the $\ket{\uparrow}$ or $\ket{\downarrow}$ state. In the strong-coupling the $m_s$ = 1/2 transitions ($Q_{\frac{1}{2}} \leftrightarrow Q_{-\frac{1}{2}}$,$D_{\frac{1}{2}} \leftrightarrow D_{-\frac{1}{2}}$) will occur at the Larmor precession frequency and the other quartet resonances ($Q_{\pm \frac{3}{2}} \leftrightarrow Q_{\pm \frac{1}{2}}$) will have a large separation from the substantial zero-field splitting. 

\subsection{Transition frequencies}

\begin{figure} [t]
\centerline{\includegraphics[width=8cm]{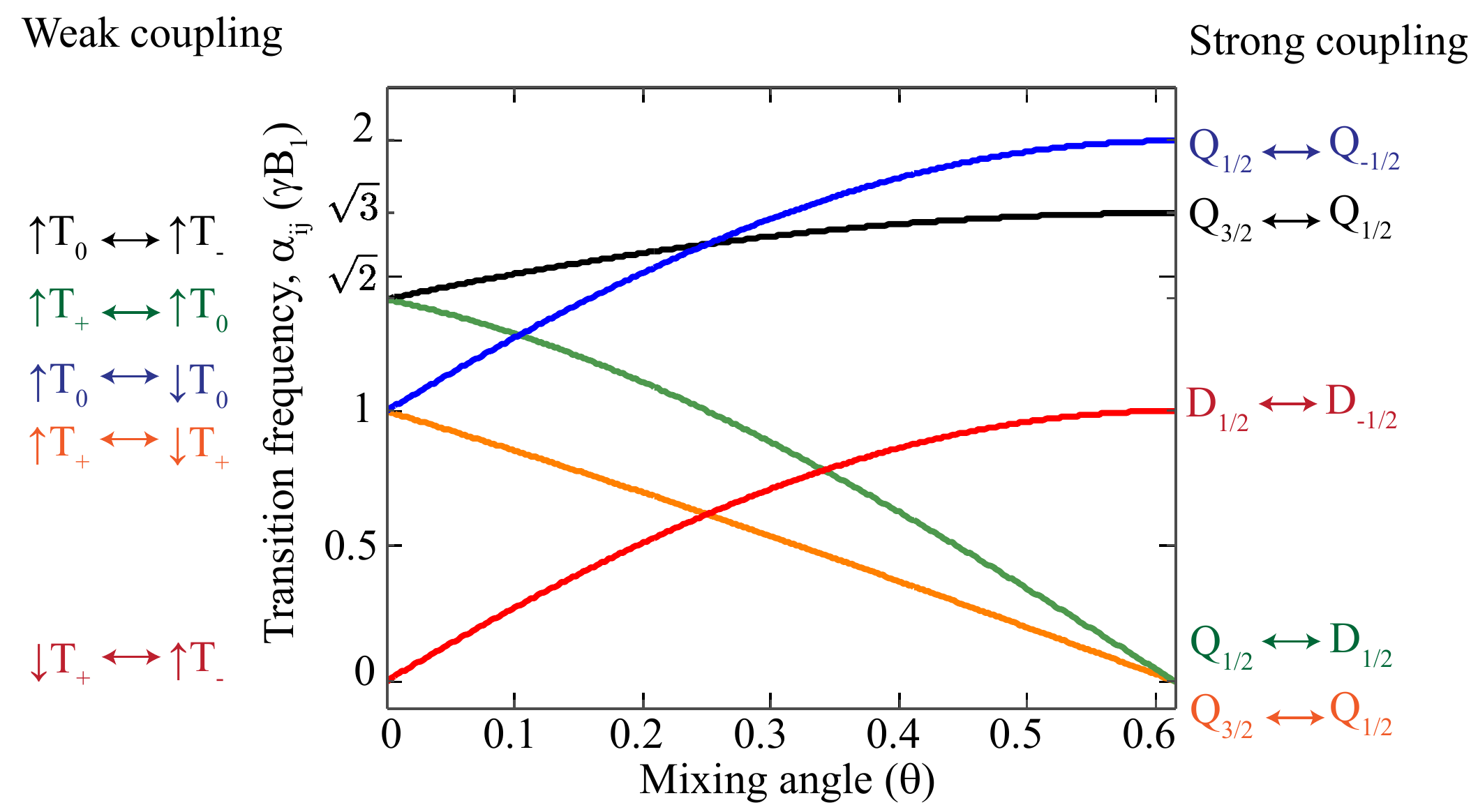}}
\caption{The transition frequencies of an exciton-polaron complex as a function of the mixing angle. In the uncoupled regime ($\theta$ = 0) the polaron will nutate at a frequency of $\gamma B_1$ and the exciton will nutate at a frequency of $\sqrt{2} \gamma B_1$. Weak coupling splits the exciton and polaron resonances and permits the $\downarrow T_+ \leftrightarrow \uparrow T_-$ transition to be weakly driven. In the strong coupling limit pure quartet and doublet states are formed. Quartet mixing will occur at frequencies of 2$\gamma B_1$ ($Q_{1/2} \leftrightarrow Q_{-1/2}$) and $\sqrt{3} \gamma B_1$ ($Q_{\pm 3/2} \leftrightarrow Q_{\pm 1/2}$). Driving can also occur within the doublet manifold at a frequency of $\gamma B_1$. Driving between the doublet and quartet states becomes arbitrarily slow in the strong coupling limit. Transition frequencies are calculated by using equation (\ref{eq:TransitionFrequency}).}
\label{fig:TransitionMixing}
\end{figure}

The frequency of the Rabi oscillations are indicative of the types of states and coupling that are present, and are calculated by solving the $2\times2$ eigenvalue problem of the relevant subspace. The frequencies of the Rabi oscillations have the form 

\begin{equation}
\Omega_{ij} = \sqrt{\alpha_{ij}^2 (\gamma B_1)^2 + (\omega - \omega_{ij})^2}
\end{equation}

Where $\alpha_{ij}$ is the frequency coefficient which may be read from the diagonal matrix entries, $\omega$ is the applied photon frequency, and $\omega_{ij}$ is the resonant (i,j) transition frequency. When a resonant excitation $(\omega = \omega_{ij})$ is applied the nutation frequency reduces to

\begin{equation}\label{eq:TransitionFrequency}
\Omega_{ij} = \alpha_{ij} \gamma B_1 
\end{equation}

The transient nutation frequency, $\alpha_{ij}$, for each possible transition is given in Table \ref{Tab:Freqs} and are shown graphically in Fig. \ref{fig:TransitionMixing} as a function of the mixing angle $\theta$ (with $\theta = \phi$). It can be seen that the nutation frequencies change distinctively as a function of mixing angle, and this can be used to differentiate triplet-exciton polaron quenching from other mechanisms such as the triplet-triplet annihilation, as well as identifying the relevant coupling regime. 

\begin{table}[h]
\caption {The transitions frequencies of an exciton-polaron complex as a function of the spin mixing angles}
\begin{ruledtabular}
\begin{tabular}{cc}
Transition & Frequency ($\gamma B_1$) \\
\hline
(1)-(2) & $\sqrt{2}\cos{\theta} + \sin{\theta}$ \\
(1)-(4) & $\cos{\theta} - \sqrt{2}\sin{\theta}$ \\
(2)-(3) & $\sqrt{2}\cos{\theta}\cos{\phi}-\sin{\phi}(\cos{\theta}+\sqrt{2}\sin{\theta})$ \\
(2)-(5) & $\cos{\phi}(\cos{\theta}+\sqrt{2}\sin{\theta})+\sqrt{2}\cos{\theta}\sin{\phi}$ \\
(3)-(4) & $\sin{\theta}(\sin{\phi}-\sqrt{2}\cos{\phi})-\sqrt{2}\cos{\theta}\sin{\phi}$ \\
(3)-(6) & $\cos{\phi}-\sqrt{2}\sin{\phi}$ \\
(4)-(5) & $-\cos{\phi}(\sin{\theta}-\sqrt{2}\cos{\theta})-\sqrt{2}\sin{\theta}\sin{\phi}$ \\
(5)-(6) & $\sin{\phi}+\sqrt{2}\cos{\phi}$
\end{tabular}
\end{ruledtabular}
\label{Tab:Freqs}
\end{table}

\subsection{Regimes}

We are able to identify several driving regimes depending on the relative energy scales present. Figure \ref{fig:TransitionMixing} shows the transition frequencies as a function of mixing angle. Additional regimes are possible when the exchange and dipolar coupling are able to cancel out on either the diagonal or off-diagonal elements of the Hamiltonian. 

\begin{description}
	\item[Uncoupled regime] 
	The simplest case, in which there is no interaction between the exciton and polaron ($J_{TD},D_{TD}$ = 0). The polaron will nutate at $\gamma B_1$ and the exciton will be split into two lines with transition frequencies of $\sqrt{2} \gamma B_1$. When the intra-exciton dipolar vanishes the exciton levels become harmonically spaced and a multi-transition analysis is required, which we do not treat in this paper. 
	\item[Effectively uncoupled regime] 
	Similar behaviour occurs in the effectively uncoupled regime, which occurs when $|D_{TD} +3J_{TD}|<<D_{T}$. The off-diagonal mixing vanishes and the physics will be identical to the weakly-coupled regime except that the transitions have a greater splitting.
	\item[Weakly coupled regime] 
	The generic weakly coupled regime corresponds to when the exciton-polaron coupling is weak ($|D_{TD}|,|J_{TD}|<<D_{T}$). The exciton and polaron resonances become weakly correlated - the $T_+(\uparrow\leftrightarrow\downarrow)$ and $T_0(\uparrow\leftrightarrow\downarrow)$ transitions will have slightly different transition frequencies, for instance, and the $T_+ \downarrow \leftrightarrow T_- \uparrow$ transition becomes weakly allowed and will provide a low frequency background.
	\item[Isoenergtic regime] 
An interesting case occurs when $D_{TD} = \frac{3}{2}J_{TD}$ and $J_{TD} << D_{T}$. The energy levels will be unshifted to first order, but there will be a small shift in the transition frequencies and the effective recombination rates.
	\item[Trionic regime] 
	The trionic regime will occur when there is strong coupling between the exciton and polaron states from exchange or dipolar coupling. In either case, the exciton and polaron become strongly entangled and form either quartet or doublet states. Driving between the quartet and doublet states ($Q_{\pm \frac{3}{2}} \leftrightarrow D_{\mp \frac{1}{2}}$, $Q_{\pm \frac{1}{2}} \leftrightarrow D_{\pm \frac{1}{2}}$) will have a vanishing small probability. Driving will also occur between the doublet states ($D_{\frac{1}{2}} \leftrightarrow D_{-\frac{1}{2}}$) at a frequency of $\gamma B_1$ and between the quartet states ($Q_{\pm \frac{3}{2}} \leftrightarrow Q_{\pm \frac{1}{2}}$, $Q_{\frac{1}{2}} \leftrightarrow Q_{-\frac{1}{2}}$) at frequencies of $\sqrt{3} \gamma B_1$ and 2$\gamma B_1$, respectively. As these transitions leave the overall quartet-doublet content unchanged, they will not produce an observable signal in the absence of spin-orbit coupling or some other additional spin-reaction pathway. 
	\item[Intermediate coupling regime] 
	Between these two extremes lies the intermediate coupling regime which involves neither pure triplet-doublet, nor quartet-doublet basis states. There will eight possible transitions with frequencies between zero and three $\gamma B_1$. Due to the large parameter space this regime encompasses a numerical approach is usually required.
	
\end{description}

\begin{figure} [t]
\centerline{\includegraphics[width=8cm]{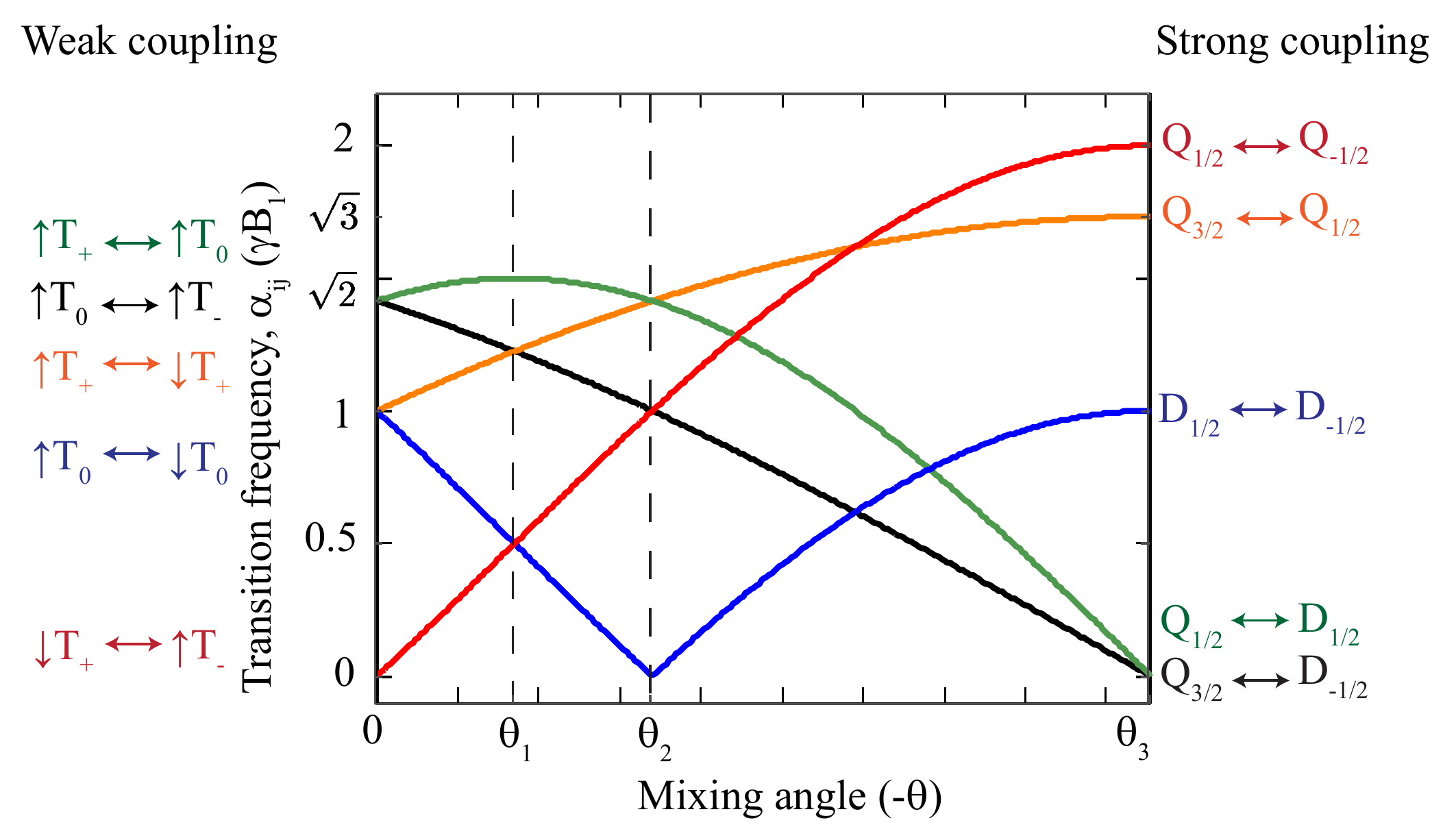}}
\caption{The transition frequencies of a bipolaron counter-ion complex, which is treated by introducing a negative mixing angle. At an angle of $\theta_1 = \frac{\cos^{-1}{\sqrt{\frac{2}{3}}}-\cos^{-1}{\sqrt{\frac{1}{3}}}}{2}$ the bipolaron is broken apart by the strong coulombic attraction from the counter-ion. If the coupling is increased further an exciton-polaron complex is formed for a mixing angle of $\theta_2 = \cos^{-1}{\sqrt{\frac{2}{3}}}-\cos^{-1}{\sqrt{\frac{1}{3}}}$. From mixing angles of $\theta_2$ to $\theta_3 = \cos^{-1}{\sqrt{\frac{1}{3}}}$ the transition frequencies of the exciton-polaron complex evolve in an identical manner to the positive mixing angle regime. Transition frequencies are calculated from equation (\ref{eq:TransitionFrequency}).}
\label{fig:Bipolaron}
\end{figure}

We have so far dealt with an exciton interacting with a polaron, which corresponds to positive exchange and dipolar coupling using our definition. A bipolaron interacting with a counter-ion can be treated by including negative exchange and dipolar coupling, a scenario advocated by Shinar\citep{shinar2012optically}. Once again we can precisely extra the spin-spin coupling from the `Rabi fingerprint'. Weaky coupling produces individually addressable states with characteristic frequencies of $\gamma B_1$ and $\sqrt{2} \gamma B_1$.  As the coupling strength is increased the bipolaron will be torn apart and an exciton-polaron will be formed. After this point the states may form a trion, proceeding in a similar manner to above. The bipolaron can be considered to be separated for a mixing angle of 

\begin{equation}
\theta = \frac{\cos^{-1}{\sqrt{\frac{2}{3}}}-\cos^{-1}{\sqrt{\frac{1}{3}}}}{2}
\end{equation}

which produces transitions of $\frac{1}{2}, \sqrt{\frac{3}{2}}$ and $\frac{3}{2}$ $\gamma B_1$. At a mixing angle of 

\begin{equation}
\theta = \cos^{-1}{\sqrt{\frac{2}{3}}}-\cos^{-1}{\sqrt{\frac{1}{3}}}
\end{equation}

an uncoupled exciton-polaron complex will be formed, with the expected transition frequencies of $\gamma B_1$ and $\sqrt{2} \gamma B_1$ for the polaron and exciton respectively. The influence of spin-spin coupling on the bipolaron is shown in Fig. \ref{fig:Bipolaron}.

\section{Half field}

It is possible to directly excite a $\Delta m_s$ = 2 spin-flip transition between the $T_-$ and $T_+$ states of an exciton at the half field resonance. 

\begin{equation}
\frac{H_1}{\hbar} = 2 \gamma B_1 [S_{e,ex}^X + S_{h,ex}^X + S_{e,P}^X] (e^{2 i \omega_0 t} + e^{-2i \omega_0 t})
\end{equation}

This transition is nominally forbidden, but becomes weakly allowed due to off-diagonal zero-field splitting which suppresses spin as a good quantum number and creates admixtures of the $T_+$ and $T_-$ states. 

To perform quantitative analysis of the half field transitions it is necessary to move to the doubly rotating frame and expand the time-averaged Hamiltonian to second order\citep{schweiger2001principles}.

\begin{equation}
\bar{H} = \frac{1}{\tau_c} \int_0^{\tau_c} H(t) dt - \frac{i}{2 \tau_c} \int_0^{\tau_c} \int_0^{t_2} [H(t_2),H(t_1)] dt_1 dt_2
\end{equation}

This produces a term proportional to

\begin{equation}
\frac{B_1 D \sin{2 \theta}}{B_0} (S_+ S_+ + S_- S_-)
\end{equation}

which is responsible for the $\Delta m_s$ = 2 transition. To obtain the correct transition probability, it is necessary to integrate over all the possible orientations with the appropriate weighting. The ensemble half-field Rabi frequency will scale with the driving field and dipolar strength in the perturbative regime.

\section{Microscopic basis for recombination}

So far we have discussed the spin-dependent recombination of exciton-polaron complexes. An obvious argument to account for the different rates is spin-conservation: The ground state is S=1/2, so we should expect the doublet states to decay faster than the quartet states, however this ignores that both are orthogonal to the ground state:

\begin{equation}
\braket{Q|0,\frac{1}{2}} = 0 
\end{equation}

\begin{equation}
\braket{D|0,\frac{1}{2}} = 0 
\end{equation}

We propose a possible mechanism in which spin conservation holds, yet allows spin angular momentum to be transferred between the exciton and polaron: A  Dexter exchange-based pathway reliant on a small asymmetry in the exciton-polaron exchange. In the canonical Hamiltonian it is tacitly assumed that the exchange between the lone polaron and each of the constituent polarons are equal. Due to the opposite polarity of the exciton polarons, one would expect there to be a small asymmetry in the two coupling strengths\citep{limes2013numerical}.

\begin{equation}
J_{TD,tot} = J_{TD} + \Delta J_{TD}
\end{equation}

Where $J_{TD,tot}$ is the total exchange coupling between the exciton and polaron, $J_{TD}$ is the sum of the two microscopic exchange interaction, and $\Delta J_{TD}$ is a small asymmetry between the electron-polaron and hole-polaron exchange coulings.

\begin{equation}
\Delta J_{TD} = S_{e,ex} \cdot S_{e,pol} - S_{h,ex} \cdot S_{e,pol}
\end{equation}

The asymmetric exciton-polaron exchange interaction ($\Delta J_{TD}$) is isotropic and will be small in comparison with the exciton exchange ($J_{T}$). The new eigenstates may be efficiently calculated analytically or numerically, but are generally cumbersome to deal with. A fruitful path is to examine the role of exciton-polaron exchange through first order perturbation theory\citep{sakurai1994modern}:

\begin{equation}
\Delta_n = \braket{n_{0}|\Delta J_{TD}|n_{0}}
\end{equation}

\begin{equation}
\ket{n} = \ket{n^{(0)}} + \Sigma_{k \neq n} \ket{k^{(0)}} \frac{V_{kn}}{E_n^{(0)} - E_k^{(0)}}
\end{equation}

Where $\Delta_n$ is the energy shift and $\ket{n}$ are the new eigenstates in terms of the uncoupled states. The resulting eigenkets are given in Table \ref{Tab:exch}

\begin{table}
\caption {The unnormalized eigenkets of a weakly coupled exciton-polaron complex when there is a small difference in the exchange coupling. The resultant mixing between singlet and triplet manifolds allows recombination of the doublet states to the singlet ground state.}
\begin{ruledtabular}
\begin{tabular}{cc}
  No Exchange & Exchange \\
\hline
  $\uparrow T_+$ & $\uparrow T_+$ \\
  $\uparrow S$ & $\uparrow S$ + $\frac{\Delta J_{TD}}{J+\frac{D}{6}} T_0$ - $\frac{\sqrt{2} \Delta J_{TD}}{J - \frac{D}{6}} \downarrow T_+$ \\
  $\uparrow T_0$ & $\uparrow T_0$ - $\frac{\Delta J_{TD}}{J+\frac{D}{6}} S$ \\
  $\uparrow T_-$ & $\uparrow T_-$ - $\frac{\sqrt{2} \Delta J_{TD}}{J - \frac{D}{6}} \downarrow S$  \\
  $\downarrow T_+$ & $\downarrow T_+$ + $\frac{\sqrt{2} \Delta J_{TD}}{J - \frac{D}{6}} \uparrow T_-$ \\
  $\downarrow S$ & $\uparrow S$ + $\frac{\Delta J_{TD}}{2(J-\frac{D}{3})} \uparrow T_0$ + $\frac{\Delta J_{TD}}{2(J-\frac{D}{6})} \downarrow T_+$ \\
  $\downarrow T_0$ & $\downarrow T_0$ + $\frac{\Delta J_{TD}}{J + \frac{D}{3}} \downarrow S$ \\
  $\downarrow T_-$ & $\downarrow T_-$ \\
\end{tabular}
\end{ruledtabular}
\end{table}\label{Tab:exch}

In this mechanism the exciton-polaron recombination occurs due to the doublet states becoming mixed with the singlet exciton. The rate of the reaction will depend on the recombination rate of the singlet exciton and the intra-exciton and exciton-polaron exchange strengths.

\begin{equation}
\braket{S|D} \approx \frac{\Delta J_{TD}}{J}
\end{equation}
\begin{equation}
R_{exciton-polaron} = R_{singlet} \frac{\Delta J_{TD}}{J}
\end{equation}

Since the singlet admixture will be proportional to the doublet content it provide the same dynamics as a process mediated directly via doublet content. A similar argument can be made involving recombination due to a slightly asymmetric dipolar coupling between the exciton and polaron.

\section{Conclusion}

Triplet-exciton polaron quenching plays an important role in the large magnetic field effects observed in organic semiconductors\citep{swanson1990optically,janssen2013tuning}. We have developed a general time-domain theory which quantitatively describes the changes in samples conductivity due to this mechanism in pulsed Electrically Detected Magnetic Resonance. Population mixing due to resonant excitation causes a net change in the exciton-charge reaction rate, and hence the free charge carrier density. 

In particular, we have derived transition frequencies and resonance positions for an exciton-polaron complex with arbitrary coupling. Our modeling indicates that there is a clearly discernible transition between an uncoupled exciton-polaron state and a strongly-bound trion. These situations may be differentiated through their transition frequencies and resonance positions. Our formalism provides a general basis for quantitative analysis of the triplet-exciton polaron process through pEDMR. In the future multi-pulse schemes such as the Hahn echo and inversion recovery could be used to investigate relaxation processes directly.

\section{Acknowledgments}

This work was supported by the Australian Research Council (ARC) (DP120102888 and CE110001027). TLK is supported by the Australian Renewable Energy Agency (ARENA)(7-S009). DRM acknowledges an ARC Future Fellowship (FT130100214).

\appendix
\section{Hamiltonian}\label{HamiltonianAppendix}
The expanded exciton-polaron Hamiltonian defined is equation (\ref{HamText}) is

\begin{widetext}
 \begin{equation}
      \begin{bmatrix}
E_1 & 0 & 0 & 0 & 0 & 0 \\
0 & \frac{\omega_s}{2} + \frac{\omega_D}{2} - \frac{2 D_T}{3} & 0 & - \frac{\sqrt{2} D_{TD}}{6} - \frac{J}{\sqrt{2}} & 0 & 0 \\
0 & 0 & \frac{\omega_s}{2} + \frac{D_T}{3} - \frac{D_{TD}}{3} + \frac{J}{2} + \frac{\omega_D}{2} - \omega_T & 0 & - \frac{\sqrt{2} D_{TD}}{6} - \frac{J}{\sqrt{2}} & 0 \\
0 & -\frac{\sqrt{2} D_{TD}}{6} - \frac{J}{\sqrt{2}} & 0 & -\frac{\omega_s}{2} + \frac{D_T}{3} - \frac{D_{TD}}{3} + \frac{J}{2} - \frac{\omega_D}{2} + \omega_T & 0 & 0 \\
0 & 0 & -\frac{\sqrt{2} D_{TD}}{6} - \frac{J}{\sqrt{2}} & 0 & -\frac{\omega_s}{2}-\frac{2 D_T}{3} - \frac{\omega_D}{2} & 0 \\
0 & 0 & 0 & 0 & 0 & E_6
      \end{bmatrix}
    \end{equation}
    \normalsize
    \end{widetext}
    
Where $E_1 = \frac{3 \omega_s}{2} + \frac{D_T}{3} + \frac{D_{TD}}{3} - \frac{J}{2} + \frac{\omega_D}{2} + \omega_T$ and $E_6 = -\frac{3 \omega_s}{2} + \frac{D_T}{3} + \frac{D_{TD}}{3} - \frac{J}{2} - \frac{\omega_D}{2} - \omega_T$.

The static Hamiltonian ($H_0$) may be transformed to the energy eigenbasis using the unitary matrix from equation (\ref{eq:unitarytrans}). 

\begin{equation}
E_1 = \frac{3 \omega_s}{2} + \frac{D_T}{3} + \frac{D_{TD}}{3} - \frac{J}{2} + \frac{\omega_D}{2} + \omega_T 
\end{equation}

\begin{multline}
E_2 = \frac{\omega_s}{2} + \frac{J}{4} - \frac{D_{TD}}{6} - \frac{D_T}{6} - \frac{\omega_T}{2} \\ 
- (\frac{1}{3}(6D_T-2D_{TD}+3J)^2+\frac{8}{3}(D_{TD}+3J)^2)
\end{multline}

\begin{multline}
E_3 = \frac{\omega_s}{2} + \frac{J}{4} - \frac{D_{TD}}{6} - \frac{D_T}{6} - \frac{\omega_T}{2} \\
+ (\frac{1}{3}(6D_T-2D_{TD}+3J)^2+\frac{8}{3}(D_{TD}+3J)^2)
\end{multline}

\begin{multline}
E_4 = -\frac{\omega_s}{2} + \frac{J}{4} - \frac{D_{TD}}{6} - \frac{D_T}{6} + \frac{\omega_T}{2} \\
+ (\frac{1}{3}(6D_T-2D_{TD}+3J)^2+\frac{8}{3}(D_{TD}+3J)^2) 
\end{multline}

\begin{multline}
E_5 = -\frac{\omega_s}{2} + \frac{J}{4} - \frac{D_{TD}}{6} - \frac{D_T}{6} + \frac{\omega_T}{2} \\
- (\frac{1}{3}(6D_T-2D_{TD}+3J)^2+\frac{8}{3}(D_{TD}+3J)^2) 
\end{multline}

\begin{equation}\label{eq:Energies}
E_6 = -\frac{3 \omega_s}{2} + \frac{D_T}{3} + \frac{D_{TD}}{3} - \frac{J}{2} - \frac{\omega_D}{2} - \omega_T
\end{equation}

The perturbative $H_1$ driving terms are strictly off-diagonal elements, which mix the eigenstates, and determine the transition frequencies. 

\begin{multline}
\\
B_{12} = B_{21} = \frac{B_1}{2}(\sin{\theta}+\sqrt{2}\cos{\theta}) \\
B_{14} = B_{41} = \frac{B_1}{2}(\cos{\theta}-\sqrt{2}\sin{\theta}) \\
B_{23} = B_{32} =  -\frac{B_1}{2}(\cos{\theta}\sin{\phi}-\sqrt{2}\cos{\theta}\cos{\phi}+\sqrt{2}\sin{\theta}\sin{\phi}) \\
B_{25} = B_{52} = \frac{B_1}{2} (\cos{\theta}\cos{\phi} + \sqrt{2}\cos{\theta}\sin{\phi} + \sqrt{2}\sin{\theta}\cos{\phi}) \\
B_{34} = B_{43} = -\frac{B_1}{2}(\sqrt{2}\cos{\theta}\sin{\phi}-\sin{\theta}\sin{\phi}+\sqrt{2}\cos{\phi}\sin{\theta})\\
B_{36} = B_{63} = \frac{B_1}{2} (\cos{\phi} - \sqrt{2} \sin{\phi}) \\
B_{45} = B_{54} = -\frac{B_1}{2} (\cos{\theta}\cos{\phi} - \sqrt{2}\cos{\theta}\sin{\phi} + \sqrt{2}\sin{\theta}\sin{\phi}) \\
B_{56} = B_{65} = \frac{B_1}{2}(\sin{\phi} + \sqrt{2} \cos{\phi})) \\
\end{multline}

Removing the Zeeman contribution from the static magnetic field ($B_0$) and substituting into equation (\ref{mat:Ham}) provides a general description of the coherent dynamics of an exciton-polaron complex.

\section{Resonance Position}

The resonance positions are given by the energy differences of the eigenstates.

\begin{equation}
\omega_{ij} = E_i - E_j
\end{equation}

These are calculated explicitly from equation (\ref{eq:Energies}) and given in Table (\ref{Tab:ResPosition}). 

\begin{table}
\caption {The resonance positions for an exciton-polaron with arbitrary spin-spin coupling, calculated from equation (\ref{eq:Energies}). In the table below X = $(\frac{1}{3}(6D_T-2D_{TD}+3J)^2+\frac{8}{3}(D_{TD}+3J)^2)$.}
\begin{ruledtabular}
\begin{tabular}{cc}
Transition & $\Delta$ E \\
\hline
(1)-(2) & $\frac{D_T}{2} + \frac{D_{TD}}{2} - \frac{3J}{4} + \frac{\omega_D}{2} + \frac{\omega_T}{2}$ + X \\
(1)-(4) & $\frac{D_T}{2} + \frac{D_{TD}}{2} - \frac{3J}{4} + \frac{\omega_D}{2} + \frac{\omega_T}{2}$ - X \\
(2)-(3) & $\omega_T$ - 2X \\
(2)-(5) & $\omega_T$ \\
(3)-(4) & $\omega_T$ \\
(3)-(6) & $\frac{3J}{4}-\frac{D_{TD}}{2}-\frac{D_T}{2}+\frac{\omega_D}{2}+\frac{\omega_T}{2}$ + X \\
(4)-(5) & $\omega_T$ + 2X \\
(5)-(6) & $\frac{3J}{4} - \frac{D_{TD}}{2} - \frac{D_T}{2} + \frac{\omega_T}{2}$ - X 
\end{tabular}
\end{ruledtabular}
\end{table}\label{Tab:ResPosition}

All of the transitions have a resonance position which depends on the spin coupling term `X', which indicates that all of the transitions will become very broad as the exciton-polaron coupling becomes large. 

\section{Microscopic basis for recombination}

The expanded (anti-symmetric) exchange matrix is 

\begin{equation}
    \Delta J_{TD} = 
      \begin{bmatrix}
0 & 0 & 0 & 0 & 0 & 0 & 0 & 0 \\
0 & 0 & \frac{\Delta J}{2} & 0 & -\frac{\Delta J}{\sqrt{2}} & 0 & 0 & 0 \\
0 & \frac{\Delta J}{2} & 0 & 0 & 0 & 0 & 0 & 0 \\
0 & 0 & 0 & 0 & 0 & \frac{\Delta J}{\sqrt{2}} & 0 & 0 \\
0 & -\frac{\Delta J}{\sqrt{2}} & 0 & 0 & 0 & 0 & 0 & 0 \\
0 & 0 & 0 & \frac{\Delta J}{\sqrt{2}} & 0 & 0 & -\frac{\Delta J}{2} & 0 \\
0 & 0 & 0 & 0 & 0 & -\frac{\Delta J}{2} & 0 & 0 \\
0 & 0 & 0 & 0 & 0 & 0 & 0 & 0
      \end{bmatrix}
    \end{equation}
    
Where the coupling term is written in the expanded singlet/triplet-doublet basis. All of the matrix lie on off-diagonal elements, which means the first order effect will be to slightly admix the eigenstates, but not change the energies. This term therefore allows the canonical triplet-doublet states to project onto the singlet-doublet states. A similar argument can be made for the exciton-polaron dipolar coupling term. 

\end{document}